\documentclass[preprint,prd,amsmath,amssymb,floatfix,nofootinbib]{revtex4}
\usepackage[colorlinks=true,urlcolor=black]{hyperref}

\hypersetup{backref = true,pagebackref = true,hyperindex = true, colorlinks = true,
  breaklinks = true, urlcolor = blue, linkcolor = blue, bookmarks = true,
  bookmarksopen = true, citecolor=red}

\usepackage{bm}
\usepackage{color}
\usepackage{dcolumn}
\usepackage{graphics}
\usepackage{graphicx}
\usepackage{subfigure}
\usepackage{slashed}
\usepackage{placeins}

\usepackage{pdfpages}
\usepackage{eso-pic}
\usepackage{everyshi}
\usepackage{pdflscape}

\newcommand{\bea}{\begin{eqnarray}}
\newcommand{\eea}{\end{eqnarray}}
\newcommand{\be}{\begin{equation}}
\newcommand{\ee}{\end{equation}}

\newcommand{\ar}{a_s}

\begin{document}

\title{Bjorken sum rule with analytic coupling
}
\author{I.R. Gabdrakhmanov$^{1}$, N.A Gramotkov$^{1,2}$, A.V.~Kotikov$^{1}$,  O.V.~Teryaev$^{1}$, D.A. Volkova$^{1,3}$  and I.A.~Zemlyakov$^{1,4}$}
\affiliation{
  $^1$Bogoliubov Laboratory of Theoretical Physics,
  Joint Institute for Nuclear Research, 141980 Dubna, Russia;\\
$^2$Moscow State University, 119991, Moscow, Russia\\
  $^3$Dubna State University,
  141980 Dubna, Moscow Region, Russia;\\
  $^4$Tomsk State University,
  634010 Tomsk,
  Russia}


\begin{abstract}

  We found good agreement between the experimental data obtained for
  the polarized Bjorken sum rule 
  and the predictions of analytic QCD, as well as a strong
  difference between these data and the results obtained in the framework of perturbative QCD. To satisfy the limit of photoproduction and take into account
  Gerasimov-Drell-Hearn and Burkhardt-Cottingham sum rules, we
  develope new representation of the perturbative part of the polarized Bjorken sum rule.

\end{abstract}

\maketitle

\section{Introduction}

The polarized
Bjorken sum rule (BSR) 
$\Gamma^{p-n}_1(Q^2)$ \cite{Bjorken:1966jh}, i.e. the difference
in the first (Mellin) moments of spin-dependent structure functions (SFs) of proton and neutron, is a very important
 space-like QCD observable \cite{Deur:2018roz,Kuhn:2008sy}.
 Its isovector nature simplifies the theoretical description 
 within the framework of perturbative QCD (pQCD) in
 comparison with the corresponding SF integrals for each nucleon.
 Experimental results for this quantity obtained in polarized deep inelastic scattering (DIS) are currently available
 in a ruther wide range of spacelike squared momenta $Q^2$: 0.021 GeV$^2\leq Q^2 <$5 GeV$^2$
\cite{Deur:2021klh,E143:1998hbs,SpinMuon:1993gcv,COMPASS:2005xxc,HERMES:1997hjr,Deur:2004ti,ResonanceSpinStructure:2008ceg}.
In particular, the most recent experimental results \cite{Deur:2021klh} with significantly reduced statistical
uncertainty make BSR an attractive value for testing various generalizations of pQCD at low $Q^2$ values:
$Q^2\leq 1$GeV$^2$.

Theoretically, pQCD (with the operator product extension (OPE)) \textcolor{green}{ in }the $\overline{MS}$ scheme, was a common approach
to describing such quantities.
This approach, however, has a theoretical disadvantage, which is that the strong coupling constant ({\it couplant})
$\alpha_s(Q^2)$ has Landau singularities for small values of $Q^2$: $Q^2\leq 0.1$GeV$^2$, which makes it inconvenient
to estimate spacelike observable quantities, such as BSR, at small values of $Q^2$.
In recent years, the extension of the QCD couplant without
Landau singularity for low $Q^2$  called 
(fractional) analytical perturbation theory [(F)APT)]
\cite{ShS,BMS1,Bakulev:2006ex} (or the
minimal analytic (MA) theory, as discussed in \cite{Cvetic:2008bn}), was
applied to compare theoretical OPE expression and experimental BSR data
\cite{Pasechnik:2008th,Khandramai:2011zd,Ayala:2017uzx,Ayala:2018ulm,Gabdrakhmanov:2023rjt}
(see also other recent BSR studies in \cite{Kotlorz:2018bxp,Ayala:2023wpy}).
Following \cite{Cvetic:2006mk}, we introduce here the derivatives (in the $k$-order of perturbation theory (PT))
\be
\tilde{a}^{(k)}_{n+1}(Q^2)=\frac{(-1)^n}{n!} \, \frac{d^n a^{(k)}_s(Q^2)}{(dL)^n},~~a^{(k)}_s(Q^2)=\frac{\beta_0 \alpha^{(k)}_s(Q^2)}{4\pi}=\beta_0\,\overline{a}^{(k)}_s(Q^2),
\label{tan+1}
\ee
which
play a key role for the construction of analytic QCD (but still have
Landau pole). 
Hereafter $\beta_0$ is the first coefficient of the QCD $\beta$-function:
\be
\beta(\overline{a}^{(k)}_s)=-{\left(\overline{a}^{(k)}_s\right)}^2 \bigl(\beta_0 + \sum_{i=1}^k \beta_i {\left(\overline{a}^{(k)}_s\right)}^i\bigr),
\label{bQCD}
\ee
where $\beta_i$ are known up to $k=4$ \cite{Baikov:2008jh}.

The series of derivatives $\tilde{a}_{n}(Q^2)$
can be used as an analogue of $\ar$-powers series, as it was numerically tested in \cite{Kotikov:2022swl}). Although each derivative reduces the $\ar$ power, on the other hand it produces an
additional $\beta$-function and, consequently, additional $ \ar^2$
factor. 
  According to the definition (\ref{tan+1}), in LO the expressions for  $\tilde{a}_{n}(Q^2)$ and $\ar^{n}$ exactly match. Beyond LO, there is one-to-one correspondence between
$\tilde{a}_{n}(Q^2)$ and $\ar^{n}$, established in \cite{Cvetic:2006mk,Cvetic:2010di} and extended to the
fractional case  in Ref. \cite{GCAK}. 

In this paper, we apply the inverse logarithmic expansion of the MA \textcolor{green}{couplant}, recently obtained in
\cite{Kotikov:2022sos,Kotikov:2023meh} for any PT order (for a brief introduction, see \cite{Kotikov:2022vnx}).
This approach is very convenient, since for LO the MA couplants have simple representations (see \cite{BMS1}) and
beyond LO the MA couplants are very close to LO ones,
especially for $Q^2 \to \infty$ and $Q^2 \to 0$, where the differences between MA couplants of different PT orders
become insignificant.
Moreover, for $Q^2 \to \infty$ and $Q^2 \to 0$ the (fractional) derivatives of the MA couplants with $n\geq 2$ tend
to zero, and therefore only the first
term in perturbative expansions makes a valuable contribution.
Along with that, the new modification of BSR allows us to make the derivative of its PT term finite in the IR limit and be in agreement with Gerasimov-Drell-Hearn and Burkhardt-Cottingham sum rules.

\section{Bjorken sum rule}

The polarized BSR
is defined as the difference between the proton and neutron polarized SFs,
integrated over the entire interval $x$
\be
\Gamma_1^{p-n}(Q^2)=\int_0^1 \, dx\, \bigl[g_1^{p}(x,Q^2)-g_1^{n}(x,Q^2)\bigr].
\label{Gpn} 
\ee

Theoretically, the quantity can be written in the OPE
form
(see Ref. \cite{Shuryak:1981pi,Balitsky:1989jb})
\be
\Gamma_1^{p-n}(Q^2)=
\frac{g_A}{6} \, \bigl(1-D_{\rm BS}(Q^2)\bigr) + \sum_{i=2}^{\infty} \frac{\mu_{2i}(Q^2)}{Q^{2i-2}} \, ,
\label{Gpn.OPE} 
\ee
where $g_A$=1.2762 $\pm$ 0.0005 \cite{PDG20} is
the nucleon axial charge, $(1-D_{BS}(Q^2))$ is the leading-twist (or twist-two)
contribution, and $\mu_{2i}/Q^{2i-2}$ $(i\geq 1)$ are the higher-twist (HT)
contributions.\footnote{Below, in our analysis, the so-called elastic contribution will always be excluded.}

Since we plan to consider in particular very small $Q^2$ values here,
the representation (\ref{Gpn.OPE}) of the HT
a number of infinite terms.
To avoid that, it is preferable to use 
the so-called "massive" twist-four representation, which includes a part of the HT
contributions of (\ref{Gpn.OPE}) (see Refs. \cite{Teryaev:2013qba,Gabdrakhmanov:2017dvg}):
\footnote{Note that Ref. \cite{Gabdrakhmanov:2017dvg} also contains a more complicated form for the "massive" twist-four
  part. It was included in our previous analysis in \cite{Gabdrakhmanov:2023rjt}, but will not be considered here.}
\be
\Gamma_1^{p-n}(Q^2)=
\frac{g_A}{6} \, \bigl(1-D_{\rm BS}(Q^2)\bigr) +\frac{\hat{\mu}_4 M^2}{Q^{2}+M^2} \, ,
\label{Gpn.mOPE} 
\ee
where the values of $\hat{\mu}_4$ and $M^2$ have been fitted in Refs. \cite{Ayala:2017uzx,Ayala:2018ulm}
in the different analytic QCD models.

In the case of MA QCD, from \cite{Ayala:2018ulm} one can see that in (\ref{Gpn.mOPE})
\be
M^2=0.439 \pm 0.012 \pm 0.463
~~\hat{\mu}_{\rm{MA},4}
=-0.173 \pm 0.002\pm 0.666\,,
\label{M,mu} 
\ee
where
the statistical (small) and systematic (large)
uncertainties are presented.

Up to the $k$-th PT order, the twist-two
part has the form
\be
D^{(1)}_{\rm BS}(Q^2)=\frac{4}{\beta_0} \, a^{(1)}_s,~~D^{(k\geq2)}_{\rm BS}(Q^2)=\frac{4}{\beta_0} \, a^{(k)}_s\left(1+\sum_{m=1}^{k-1} d_m \bigl(a^{(k)}_s\bigr)^m
\right)\,,
\label{DBS} 
\ee
where $d_1$, $d_2$ and $d_3$ are known from exact calculations (see, for example, \cite{Chen:2006tw}).
The exact $d_4$ value is not known, but it was  estimated  in Ref. \cite{Ayala:2022mgz}.

Converting the couplant powers into its derivatives, we have
\be
D^{(1)}_{\rm BS}(Q^2)=\frac{4}{\beta_0} \, \tilde{a}^{(1)}_1,~~D^{(k\geq2)}_{\rm BS}(Q^2)=
\frac{4}{\beta_0} \, \left(\tilde{a}^{(k)}_{1}+\sum_{m=2}^k\tilde{d}_{m-1}\tilde{a}^{(k)}_{m}
\right),
\label{DBS.1} 
\ee
where
\bea
&&\tilde{d}_1=d_1,~~\tilde{d}_2=d_2-b_1d_1,~~\tilde{d}_3=d_3-\frac{5}{2}b_1d_2-\bigl(b_2-\frac{5}{2}b^2_1\bigr)\,d_1,\nonumber \\
&&\tilde{d}_4=d_4-\frac{13}{3}b_1d_3 -\bigl(3b_2-\frac{28}{3}b^2_1\bigr)\,d_2-\bigl(b_3-\frac{22}{3}b_1b_2+\frac{28}{3}b^3_1\bigr)\,d_1
\label{tdi} 
\eea
and $b_i=\beta_i/\beta_0^{i+1}$.

For the case of 3 active quark flavors ($f=3$), which is accepted in this paper, we have
\footnote{
  The coefficients $\beta_i$ $(i\geq 0)$ of the QCD $\beta$-function (\ref{bQCD})
  and, consequently, the couplant $\alpha_s(Q^2)$ itself depend on the number $f$ of active quark flavors, and each new
  quark enters/leaves
  the game at a certain threshold $Q^2_f$ according to \cite{Chetyrkin:2005ia}. The
  corresponding QCD parameters  $\Lambda^{(f)}$ in N$^i$LO of PT can be found
  in Ref. \cite{Chen:2021tjz}.}
\bea
&&d_1=1.59,~~d_2=3.99,~~d_3=15.42~~d_4=63.76, \nonumber \\ 
&&\tilde{d}_1=1.59,~~\tilde{d}_2=2.73,
~~\tilde{d}_3=8.61,~~\tilde{d}_4=21.52 \, ,
\label{td123} 
\eea
i.e., the coefficients in the series of derivatives are slightly smaller.

In MA QCD, the results (\ref{Gpn.mOPE}) become as follows
\be
\Gamma_{\rm{MA},1}^{p-n}(Q^2)=
\frac{g_A}{6} \, \bigl(1-D_{\rm{MA,BS}}(Q^2)\bigr) +\frac{\hat{\mu}_{\rm{MA},4}M^2}{Q^{2}+M^2},~~
\,,
\label{Gpn.MA} 
\ee
where the perturbative part $D_{\rm{BS,MA}}(Q^2)$
takes
the same form, however, with analytic couplant $\tilde{A}^{(k)}_{\rm MA,\nu}$ (the corresponding expressions are taken from \cite{Kotikov:2022sos})
\be
D^{(1)}_{\rm MA,BS}(Q^2)=\frac{4}{\beta_0} \, A_{\rm MA}^{(1)},~~
D^{k\geq2}_{\rm{MA,BS}}(Q^2) =\frac{4}{\beta_0} \, \Bigl(A^{(1)}_{\rm MA}
+ \sum_{m=2}^{k} \, \tilde{d}_{m-1} \, \tilde{A}^{(k)}_{\rm MA,\nu=m} \Bigr)\,.
\label{DBS.ma} 
\ee

\section{Results}

\begin{table}[t]
\begin{center}
\begin{tabular}{|c|c|c|c|}
\hline
& $M^2$ for $Q^2 \leq 5$ GeV$^2$ & $\hat{\mu}_{\rm{MA},4}$  for $Q^2 \leq 5$ GeV$^2$& $\chi^2/({\rm d.o.f.})$ for $Q^2 \leq 5$ GeV$^2$ \\
& (for $Q^2 \leq 0.6$ GeV$^2$) & (for $Q^2 \leq 0.6$ GeV$^2$) & (for $Q^2 \leq 0.6$ GeV$^2$) \\
 \hline
 LO & 0.472 $\pm$ 0.035 & -0.212 $\pm$ 0.006 & 0.667  \\
 & (1.631 $\pm$ 0.301) & (-0.166 $\pm$ 0.001) & (0.789)  \\
 \hline
 NLO & 0.414 $\pm$ 0.035 & -0.206 $\pm$ 0.008 & 0.728  \\
& (1.545 $\pm$ 0.287) & (-0.155 $\pm$ 0.001) & (0.757)  \\
 \hline
 N$^2$LO & 0.397 $\pm$ 0.034 & -0.208$\pm$ 0.008 & 0.746  \\
 & (1.417 $\pm$ 0.241) & (-0.156 $\pm$ 0.002) & (0.728)  \\
 \hline
 N$^3$LO & 0.394 $\pm$ 0.034 & -0.209 $\pm$ 0.008 & 0.754  \\
  & (1.429 $\pm$ 0.248) & (-0.157 $\pm$ 0.002) & (0.747)  \\
   \hline
N$^4$LO & 0.397 $\pm$ 0.035 & -0.208 $\pm$ 0.007 & 0.753  \\
 & (1.462 $\pm$ 0.259) & (-0.157 $\pm$ 0.001) & (0.754)  \\
 \hline
\end{tabular}
\end{center}
\caption{%
  The values of the fit parameters in (\ref{Gpn.MA}).
}
\label{Tab:BSR}
\end{table}

\begin{figure}[t]
\centering
\includegraphics[width=0.98\textwidth]{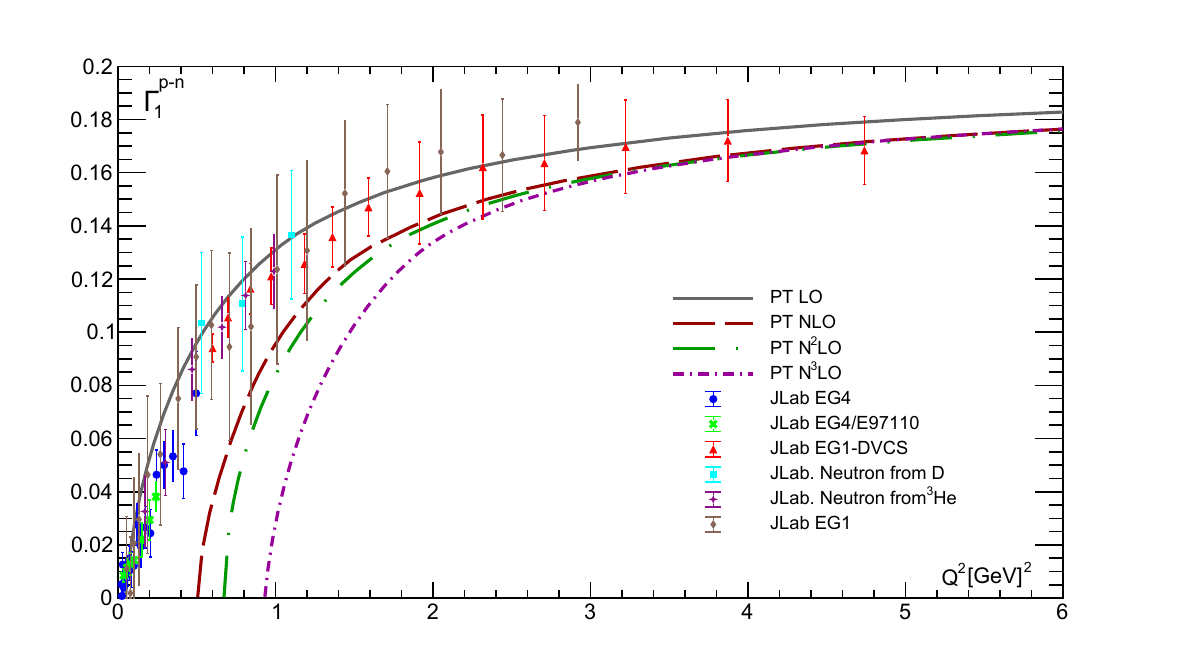}
\caption{
  \label{fig:PT}
  The results for $\Gamma_1^{p-n}(Q^2)$ in the first  four
  orders of PT.
    }
\end{figure}

\begin{figure}[t]
\centering
\includegraphics[width=0.98\textwidth]{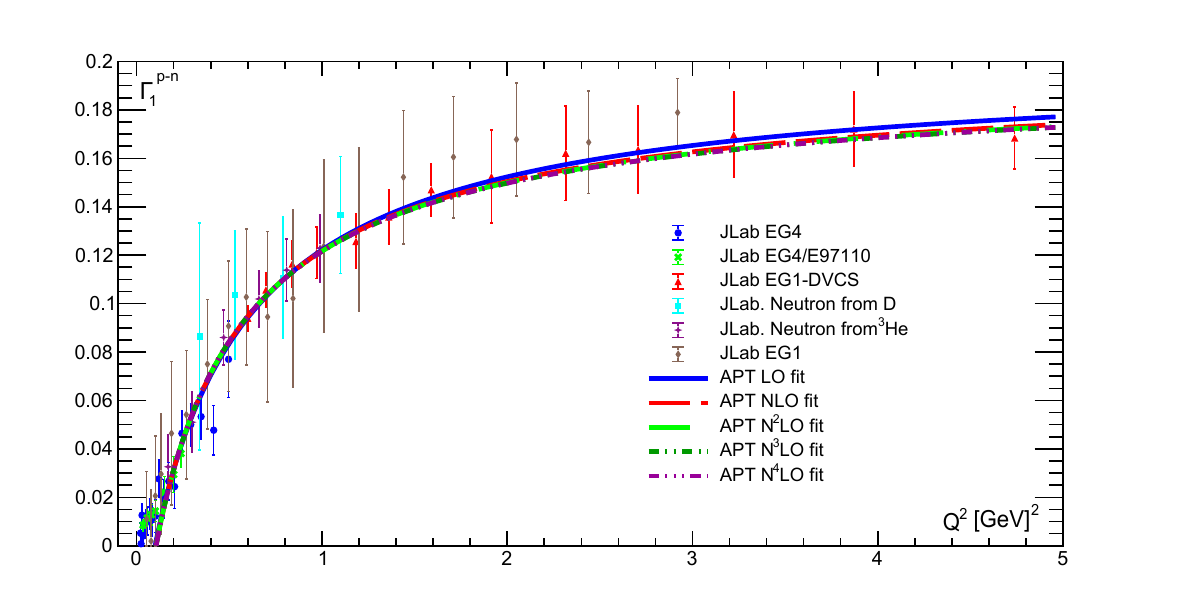}
\caption{
  \label{fig:APT}
  The results for $\Gamma_1^{p-n}(Q^2)$ in the first  four
  orders of APT.
    }
\end{figure}

\begin{figure}[t]
\centering
\includegraphics[width=0.98\textwidth]{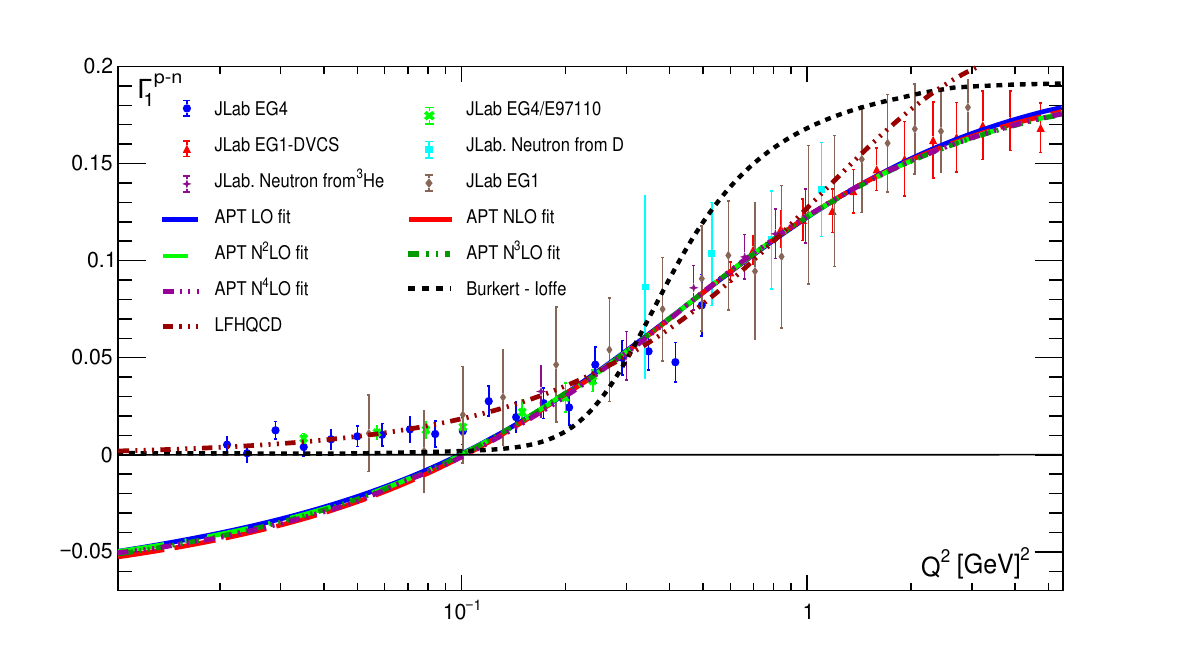}
\caption{
  \label{fig:APTHT}
  Same as in
  Fig. \ref{fig:APT}  but for $Q^2 <$0.6 GeV$^2$.
}
\end{figure}

The fitting results  of experimental data obtained only with
statistical uncertainties 
are presented in Table \ref{Tab:BSR} and shown in Figs.
\ref{fig:PT} and \ref{fig:APT}.
For the fits we use $Q^2$-independent $M^2$ and $\hat{\mu}_4$
and the two-twist part shown in Eqs. (\ref{DBS.1}), (\ref{DBS.ma}) 
for regular PT and APT, respectively.

As it can be seen in Fig. \ref{fig:PT}, with the exception of LO, the results obtained using conventional couplant are very poor.
Moreover, the
discrepancy 
in this case increases with the order of PT (see also
\cite{Pasechnik:2008th,Khandramai:2011zd,Ayala:2017uzx,Ayala:2018ulm} for similar analyses).
The LO results
describe experimental points 
  relatively well, since the value of
$\Lambda_{\rm LO}$ is quite small compared to other $\Lambda_{i}$, and
disagreement with the data begins at lower values of $Q^2$ (see Fig. 4
below).
Thus, using  the ``massive'' twist-four form (\ref{Gpn.mOPE}) does not improve these results, since with
$Q^2 \to \Lambda_i^2$ conventional couplants become singular, which leads to large and negative results for
the twist-two part (\ref{DBS}). So, as the PT order increases,
ordinary couplants become singular for ever larger $Q^2$ values, while BSR tends to negative values for ever
larger $Q^2$ values.

In contrast, our results obtained for different APT orders are practically equivalent:
the corresponding curves become indistinguishable when $Q^2$ approaches 0 and slightly different everywhere else. As can be seen in Fig. \ref{fig:APT}, the fit quality is pretty high,
which is demonstrated 
  by the values of the corresponding $\chi^2/({\rm d.o.f.})$ (see Table \ref{Tab:BSR}).

\subsection{Low  $Q^2$ values}

\begin{figure}[t]
\centering
\includegraphics[width=0.98\textwidth]{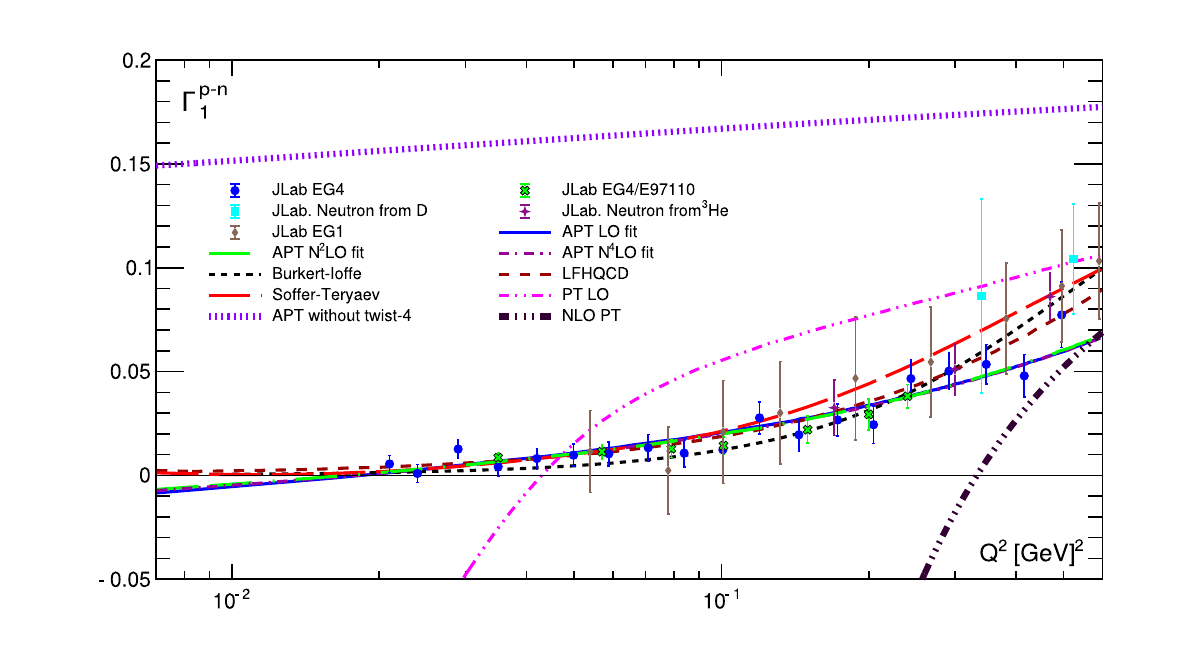}
\caption{
  \label{fig:low}
The results for $\Gamma_1^{p-n}(Q^2)$ in the first four orders of APT 
from fits of experimental data with $Q^2 <$0.6 GeV$^2$
}
\end{figure}

The full picture, however, is more complex than shown in Fig. \ref{fig:APT}.
The APT fitting curves 
become negative (see Fig. \ref{fig:APTHT}) when we
move to very low values of $Q^2$:
$Q^2 <$0.1 GeV$^2$.
So, the good quality of the fits shown in Table \ref{Tab:BSR} was obtained due to good
agreement with experimenatl data at $Q^2 >$0.2 GeV$^2$.
The picture improves significantly when we compare our result with experimental data for $Q^2~<$0.6 GeV$^2$
(see Fig. \ref{fig:low} and Ref. \cite{Gabdrakhmanov:2023rjt}).

Fig. \ref{fig:low} also shows contributions from conventional PT in
the first two orders: the LO and NLO predictions
have nothing in common
with experimental data. As we mentioned 
above, higher orders lead to even worse agreement,
and they are not shown. The purple curve emphasizes the key role of
the twist-four contribution
(see also \cite{Khandramai:2011zd}, \cite{Kataev:2005ci}
and the discussions therein).
Excluding this contribution, the value of $\Gamma_1^{p-n}(Q^2)$ is about 0.16, which is very far from the experimental data.

At $Q^2\leq 0.3$GeV$^2$, we also see the good
agreement with the phenomenological models:
LFHQCD \cite{Brodsky:2014yha} and the correct IR limit of Burkert--Ioffe model
\cite{Burkert:1992tg}.
For larger values of $Q^2$, our results are
lower than the results of phenomenological models, and for $Q^2\geq
0.5$GeV$^2$ below the experimental data.

Nevertheless, even in this case where very good agreement with experimental data with $Q^2~ <$0.6 GeV$^2$ is demonstrated,
our results for $\Gamma_{\rm{MA},1}^{p-n}(Q^2)$ take negative
unphysical values when $Q^2 <$0.02 GeV$^2$. 
The reason for this
phenomenon can be shown by considering photoproduction within APT, which is the topic of the next subsection.

\subsection{ Photoproduction }

To understand the problem $\Gamma_{\rm{MA},1}^{p-n}(Q^2\to 0)<0$,
demonstrated above, we 
consider the photoproduction case.
  In the $k$-th order of MA QCD
  \be
  A^{(k)}_{\rm MA}(Q^2=0)\equiv \tilde{A}^{(k)}_{{\rm MA},m=1}(Q^2=0)=1,~~ \tilde{A}^{(k)}_{{\rm MA},m}=0,~~\mbox{when}~~m>1
\label{Akm} 
\ee
  and, so, we have
  \be
D_{\rm MA,BS}(Q^2=0)=\frac{4}{\beta_0} 
~~\mbox{and, hence,}~~
\Gamma_{\rm{MA},1}^{p-n}(Q^2=0)=
\frac{g_A}{6} \, \bigl(1-\frac{4}{\beta_0}\bigr) +\hat{\mu}_{\rm{MA},4}
\,.
\label{Gpn.MA.Q0} 
\ee
The finitness of cross-section in the real photon limit 
leads to \cite{Teryaev:2013qba}
\be
\Gamma_{\rm{MA},1}^{p-n}(Q^2=0)=0
~~\mbox{and, thus,}~~
\hat{\mu}^{php}_{\rm{MA},4}=-\frac{g_A}{6} \, \bigl(1-\frac{4}{\beta_0}\bigr).
\label{mu.GDH} 
\ee
For 
$f=3$, we have
\be 
\hat{\mu}^{php}_{\rm{MA},4}=-0.118
~~\mbox{and, hence,}~~
|\hat{\mu}^{php}_{\rm{MA},4}|< |\hat{\mu}_{\rm{MA},4}|,
\label{mu.GDH} 
\ee
shown in (\ref{M,mu}) and in Table \ref{Tab:BSR}.

So, as can be seen from Table \ref{Tab:BSR}, the finiteness of the cross section in the real photon limit is violated in our approaches.
\footnote{
Note that the results for $\hat{\mu}_{\rm{MA},4}$ were obtained taking into account only statistical uncertainties.
When adding systematic uncertainties, the results for $\hat{\mu}^{php}_{\rm{MA},4}$ and $\hat{\mu}_{\rm{MA},4}$ are
completely consistent with each other, but the predictive power of such an analysis is small.}
This violation leads to negative values of $\Gamma_{\rm{MA},1}^{p-n}(Q^2\to 0)$.
Note that this violation is less for experimental data sets
with $Q^2\leq 0.6$GeV$^2$, where the obtained values for
$|\hat{\mu}_{\rm  MA,4}|$ are essentially less then those obtained in
the case of experimental data with $Q^2\leq 5$GeV$^2$. Smaller values of
$|\hat{\mu}_{\rm  MA,4}|$ lead to negative values of 
$\Gamma_{\rm{MA},1}^{p-n}(Q^2\to 0)$, when $Q^2\leq 0.02$GeV$^2$ (see
Fig. 4).

\subsection{ Gerasimov-Drell-Hearn and Burkhardt-Cottingham sum rules}

Now  we plan to improve this analysis by involving the result (\ref{Gpn.MA}) at low $Q^2$ values
and also taking into account the ``massive'' twist-six term, similar to the twist-four shown in Eq. (\ref{Gpn.mOPE}).

Moreover, we take into account also the Gerasimov-Drell-Hearn (GDH) and Burkhardt-Cottingham (BC) sum rules, which lead to
(see \cite{Teryaev:2013qba,Gabdrakhmanov:2017dvg,Soffer:1992ck,Pasechnik:2010fg})
  \be
  \frac{d}{dQ^2} \Gamma_{\rm{MA},1}^{p-n}(Q^2=0)= G,~~G=\frac{\mu^2_n-(\mu_p-1)^2}{8M_p^2}=0.0631\,,
  \label{GDH} 
\ee
where $\mu_n=-1.91$ and $\mu_p=2.79$ are proton and neutron magnetic moments, respectively,
and $M_p$ = 0.938 GeV is a nucleon mass. Note that the value of $G$ is small.

In agreement with the definition (\ref{tan+1}), we have that 
\be
Q^2\frac{d}{dQ^2} \tilde A_n(Q^2) \sim \tilde A_{n+1}(Q^2)\,.
  \label{An} 
\ee

Then, for $Q^2 \to 0$ we obtain at any $n$ value, that
\be
Q^2\frac{d}{dQ^2}  \tilde A_n(Q^2) \to 0\,,
  \label{AnQ0} 
\ee
but very slowly, that the derivative
\be
\frac{d}{dQ^2}  \tilde A_n(Q^2\to 0) \to \infty \,.
  \label{AnQ0} 
\ee

Thus,  after application the derivative $d/dQ^2$ for
$\Gamma_{\rm{MA},1}^{p-n}(Q^2)$ , every term in $D_{\rm MA,BS}(Q^2)$
becomes to be divergent at $Q^2 \to 0$.
To produce finitness at $Q^2 \to 0$ for the l.h.s. of (\ref{GDH}),
we can assume the relation between twist-two and twist-four terms,
that leads to the appearance of a new contribution 
\be
-\frac{g_A}{6}\,D_{\rm MA,BS}(Q^2) +  \frac{\hat{\mu}_{\rm{MA},4}M^2}{Q^{2}+M^2}\,D_{\rm MA,BS}(Q^2)\,,
  \label{DD} 
  \ee  
which can be done to be regular at $Q^2 \to 0$.

The form (\ref{DD}) suggests the following idea about a modification of $\Gamma_{\rm{MA},1}^{p-n}(Q^2)$ in (\ref{Gpn.MA}):
\be
\Gamma_{\rm{MA},1}^{p-n}(Q^2)=
\, \frac{g_{A}}{6} \, \bigl(1-D_{\rm{MA,BS}}(Q^2) \cdot \frac{Q^2}{Q^2+M_2^2}\bigr) +
\frac{\hat{\mu}_{\rm{MA},4}M_4^2}{Q^{2}+M_4^2}+\frac{\hat{\mu}_{\rm{MA},6}M_6^4}{(Q^{2}+M_6^2)^2},~~
\label{Gpn.MAn} 
\ee
where we added the ``massive'' twist-six term and introduced different masses in both higher-twist terms and into
the modification factor $Q^2/(Q^2+M_2^2)$.

The finitness of cross-section in the real photon limit  leads now to \cite{Teryaev:2013qba}
\be
\Gamma_{\rm{MA},1}^{p-n}(Q^2=0)=0=\frac{g_{A}}{6}+\hat{\mu}_{\rm{MA},4}+\hat{\mu}_{\rm{MA},6}\,
\label{Gpn.MAnQ0} 
\ee
and, thus, we have
\be
\hat{\mu}_{\rm{MA},4}+\hat{\mu}_{\rm{MA},6}=-\frac{g_{A}}{6} \approx - 0.21205
\label{Gpn.MAnQ0.1} 
\ee

\begin{table}[t]
\begin{center}
\begin{tabular}{|c|c|c|}
\hline
& $M^2$ for $Q^2 \leq 5$ GeV$^2$& $\chi^2/({\rm d.o.f.})$ for $Q^2 \leq 5$ GeV$^2$ \\
& (for $Q^2 \leq 0.6$ GeV$^2$)  & (for $Q^2 \leq 0.6$ GeV$^2$) \\
 \hline
 LO & 0.383 $\pm$ 0.014 (0.576 $\pm$ 0.046) & 0.572 (0.575) \\
 \hline
 NLO & 0.394 $\pm$ 0.013  (0.464 $\pm$ 0.039) & 0.586 (0.590) \\
 \hline
 N$^2$LO & 0.328 $\pm$ 0.014 (0.459 $\pm$ 0.038) & 0.617 (0.584) \\
 \hline
 N$^3$LO & 0.330 $\pm$ 0.014  (0.464 $\pm$ 0.039) & 0.629 (0.582) \\
   \hline
N$^4$LO & 0.331 $\pm$ 0.013  (0.465 $\pm$ 0.039) & 0.625  (0.584) \\
 \hline
\end{tabular}
\end{center}
\caption{%
  The values of the fit parameters.
}
\label{Tab:BSR1}
\end{table}

From Eq. (\ref{Gpn.MAn}) and condition (\ref{GDH}), we obtain
\be
-\frac{g_{A}}{6}\cdot \frac{D_{\rm{MA,BS}}(Q^2=0)}{M_2^2} -\frac{\hat{\mu}_{\rm{MA},4}}{M_4^2}-2\frac{\hat{\mu}_{\rm{MA},6}}{M_6^2}=G\,,
\label{Gpn.MAnQ0.2} 
\ee
where $D_{\rm{MA,BS}}(Q^2=0)=4/\beta_0$ (see Eq. (\ref{Gpn.MA.Q0})).

Using $f=3$ (i.e. $\beta_0=9$) and putting, for simplicity, $M_2=M_4=M_6=M$, we have 
\be
\hat{\mu}_{\rm{MA},4}+2\hat{\mu}_{\rm{MA},6}=-G\,M^2-\frac{2g_{A}}{3\beta_0}=-G\,M^2-\frac{2g_{A}}{27} \approx -G\,M^2- 0.0945
\label{Gpn.MAnQ0.3} 
\ee

Taking the results (\ref{Gpn.MAnQ0}) and (\ref{Gpn.MAnQ0.3}) together,
we have at the end the following results:
\bea
\hat{\mu}_{\rm{MA},6} =-G\,M^2+\frac{5g_{A}}{54}= -G\,M^2+0.1182,\nonumber \\
\hat{\mu}_{\rm{MA},4} = -\frac{g_{A}}{6} -\hat{\mu}_{\rm{MA},6}= G\,M^2-\frac{7g_{A/V}}{27}=  G\,M^2-0.3309\,.
\label{Gpn.MAnQ0.4} 
\eea

Since the value of $G$ is small, so $\hat{\mu}_{\rm{MA},4}<0$ and $\hat{\mu}_{\rm{MA},4} \approx -0.36\hat{\mu}_{\rm{MA},4}>0$.

\begin{figure}[t]
\centering
\includegraphics[width=0.98\textwidth]{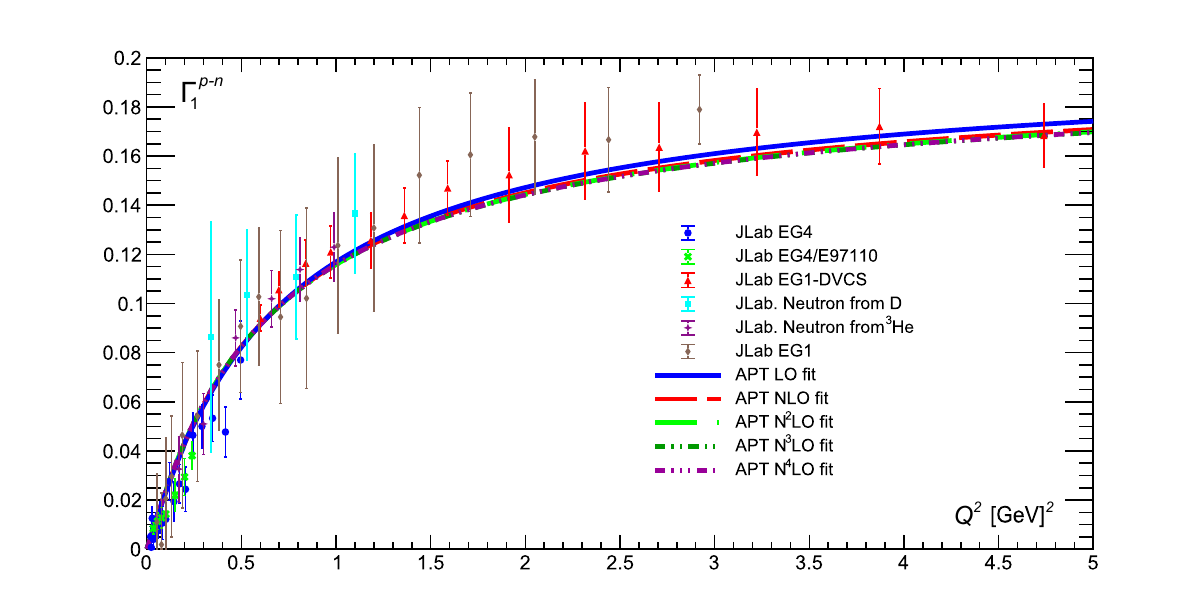}
\caption{
  \label{fig:PT1}
  The results for $\Gamma_1^{p-n}(Q^2)$ (\ref{Gpn.MAn})
  in the first  four  orders of APT. 
}
\end{figure}

\begin{figure}[t]
\centering
\includegraphics[width=0.98\textwidth]{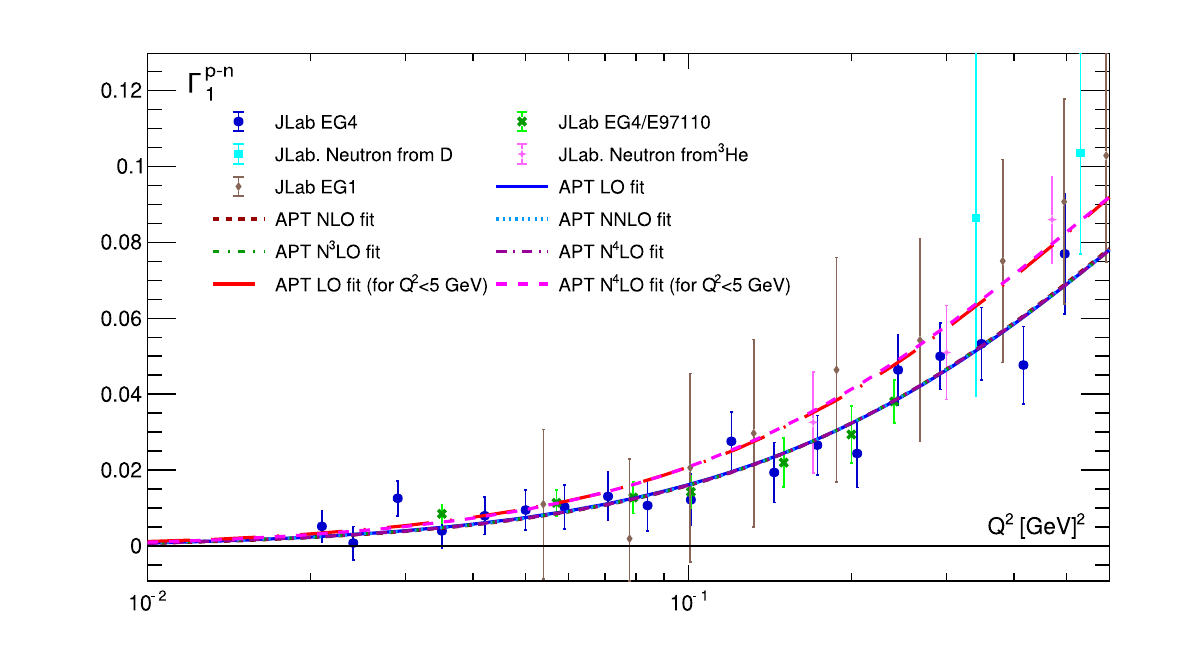}
\caption{
  \label{fig:low1}
  As in Fig. \ref{fig:PT1} but for $Q^2<$0.6 GeV$^2$
}
\end{figure}

The fitting results  of theoretical predictions based on
Eq. (\ref{Gpn.MAn}) with $\hat{\mu}_{\rm{MA},4}$ and
$\hat{\mu}_{\rm{MA},6}$ done in (\ref{Gpn.MAnQ0.4}) (i.e. with the
condition $M_2=M_4=M_6=M$),
are presented in Table \ref{Tab:BSR1} and on Figs. \ref{fig:PT1} and \ref{fig:low1}.

As one can see in Table \ref{Tab:BSR1}, the obtained results for $M^2$ are different if we
take the full data set and the
limited one with $Q^2<$0.6 GeV$^2$.
However, the difference is significantly less than it was in Table \ref{Tab:BSR}.
Moreover, the results obtained in the fits using the full
data set
and shown in Tables \ref{Tab:BSR} and \ref{Tab:BSR1} are quite similar, too.

We also see some similarities between the results shown in Figs. \ref{fig:APT}  and \ref{fig:PT1}.
The difference appears only
at small $Q^2$ values, as can be seen in Figs. \ref{fig:APTHT} and \ref{fig:low1}.

Fig. \ref{fig:low1} also shows that the results of fitting the full set of
experimental data are in better agreement with
the data at $Q^2\geq 0.55$GeV$^2$, as it should be, since these data are involved in the analyses of the full set of experimental data.\\

Of course, the low $Q^2$ modification (\ref{Gpn.MAn}) of the result (\ref{Gpn.MA}) is not unical.
There are other possibilities. One of them can be represented as
\be
\Gamma_{\rm{MA},1}^{p-n}(Q^2)=
\, \frac{g_{A/V}}{6} \, \bigl(1-D_{\rm{MA,BS}}(Q^2)\bigr) \cdot \frac{Q^2}{Q^2+M_2^2} +
\frac{\hat{\mu}_{\rm{MA},4}M_4^2}{Q^{2}+M_4^2}+\frac{\hat{\mu}_{\rm{MA},6}M_6^4}{(Q^{2}+M_6^2)^2}.~~
\label{Gpn.MAnA} 
\ee

The finitness of cross-section in the real photon limit  leads now to
\be
\Gamma_{\rm{MA},1}^{p-n}(Q^2=0)=0=
\hat{\mu}_{\rm{MA},4}+\hat{\mu}_{\rm{MA},6}\,
\label{Gpn.MAnQ0A} 
\ee
and, so, we have the relation
\be
\hat{\mu}_{\rm{MA},4}+\hat{\mu}_{\rm{MA},6}=0,~~\mbox{or}~~\hat{\mu}_{\rm{MA},4}=-\hat{\mu}_{\rm{MA},6}
\label{Gpn.MAnQ0.1A} 
\ee

From (\ref{Gpn.MAnA}) and (\ref{GDH}), we have
\be
\frac{g_{A}}{6M_2^2}\cdot \left(1- D_{\rm{MA,BS}}(Q^2=0)\right) -\frac{\hat{\mu}_{\rm{MA},4}}{M_4^2}-2\frac{\hat{\mu}_{\rm{MA},6}}{M_6^2}=-G\,.
\label{Gpn.MAnQ0.2A} 
\ee
Using $f=3$ (and, thus, $\beta_0=9$) and also
$M_2=M_4=M_6=M$, we have
\be
\hat{\mu}_{\rm{MA},4}+2\hat{\mu}_{\rm{MA},6}=-G\,M^2+\frac{g_{A/V}}{6}\cdot \left(1-\frac{4}{\beta_0}\right) =-G\,M^2+\frac{5g_{A/V}}{54} \approx -G\,M^2+0.1182
\label{Gpn.MAnQ0.3A} 
\ee

So, from (\ref{Gpn.MAnQ0.1A}) and (\ref{Gpn.MAnQ0.3A}) we obtain
\bea
\hat{\mu}_{\rm{MA},6} =-G\,M^2+\frac{5g_{A/V}}{54} \approx -G\,M^2+0.1182,\nonumber \\
\hat{\mu}_{\rm{MA},4} =
-\hat{\mu}_{\rm{MA},6}= G\,M^2-\frac{5g_{A/V}}{54} \approx G\,M^2-0.1182\,.
\label{Gpn.MAnQ0.4A} 
\eea

We note that
at the case $M_2=M_4$ the results (\ref{Gpn.MAn}) and (\ref{Gpn.MAnA}) are equal and
are related with the replacement:
\be
\hat{\mu}_{\rm{MA},4}\to \frac{g_{A/V}}{6}+\hat{\mu}_{\rm{MA},4}\,.
\label{Gpn.MAnQ0.4A1} 
\ee

So, if we use Eq.(\ref{Gpn.MAnA}) with the condition $M_2=M_4=M_6=M$
for numerical analyses, the results should be equivalent to the
results shown in Table \ref{Tab:BSR1}. We have verified
this numerically.

\section{Conclusions}

We have considered the Bjorken sum rule
$\Gamma_{1}^{p-n}(Q^2)$ in the framework of MA and
perturbative
QCD and obtained results similar to those obtained in previous studies
\cite{Pasechnik:2008th,Khandramai:2011zd,Ayala:2017uzx,Ayala:2018ulm,Gabdrakhmanov:2023rjt}
for the first 4 orders of PT.
The results based on the conventional PT do not agree with the experimental data. For some $Q^2$ values, the PT results become negative, since the
high-order corrections are large and enter the twist-two term with a minus sign.
APT in the minimal version leads to a good agreement with experimental data when we used the ``massive'' version (\ref{Gpn.MA}) for
the twist-four contributions.

Examining low $Q^2$ behaviour, we found that there is a disagreement between the results obtainded in the fits and
application of MA QCD to photoproduction. The results of fits extented to low $Q^2$ lead to the negative values for
Bjorken sum rule $\Gamma_{\rm{MA},1}^{p-n}(Q^2)$:  $\Gamma_{\rm{MA},1}^{p-n}(Q^2\to 0) <0$ that contrary to the
finitness of cross-section in the real photon limit, which leads to
$\Gamma_{\rm{MA},1}^{p-n}(Q^2\to 0) =0$. Note that
fits of experimental data at low $Q^2$ values (we used $Q^2<$ 0.6 GeV$^2$) lead to less magnitudes of negative values
for $\Gamma_{\rm{MA},1}^{p-n}(Q^2)$.

To solve the problem we considered low $Q^2$ modifications of OPE
formula for $\Gamma_{\rm{MA},1}^{p-n}(Q^2)$.
Considering
carefully one of them, Eq. (\ref{Gpn.MAn}),
we find good agreemnet with full sets of experimental data for Bjorken sum rule
$\Gamma_{\rm{MA},1}^{p-n}(Q^2)$ and also with its $Q^2 \to 0$ limit, i.e. with photoproduction. 
We see also good agreement with phenomenological modeles, especially
with LFHQCD \cite{Brodsky:2014yha}.

As the next step in our research, we plan to add to our analysis the
heavy-quark contibution to $\Gamma_{\rm{MA},1}^{p-n}(Q^2)$.
It was calculated in a closed form in Ref. \cite{Blumlein:2016xcy}.
It is suppressed by the factor $a_s^2$, but contains a contribution of
$\sim\ln(1/Q^2)$ at low $Q^2$ values and should be important there.


{\bf Acknowledgments}~~Authors are grateful to Alexandre P. Deur for
information about new experimental data in Ref. \cite{Deur:2021klh} and discussions.
Authors thank Andrei Kataev and
Nikolai Nikolaev for careful discussions.
This work was supported in part by the Foundation for the Advancement of Theoretical
Physics and Mathematics “BASIS”.



\end{document}